\documentclass[12pt,a4paper]{article}
\title{On Distribution of Zeros of Some Entire Functions}
\author{H.I.Ahmadov \footnote{E-mail: hikmatahmadov@yahoo.com} \\
Chair of Mathematical Physics, Faculty of Applied \\
Mathermatics and Cybernetics, Baku State University \\
Z.Khalilov str.,23, 370148 Baku, Azerbaijan}

\input tcilatex
\date{}

\begin{document}

\large

\maketitle
\begin{abstract}
In this paper we investigate distribution of zeros for once quasipolynom and
obtain exactly lower-bound for their modulus
\end{abstract}
\noindent
As is known [1], [2] in connection with the investigation of completenecess
of a system of eigen and adjoint elements of definite class of spectral
problems in general not being regular by Tamarkin-Rasulov [5], there arises
the necessity of studying properties of entire analytical functions of the
form

\begin{equation}
f_k(\lambda)= e^\lambda +A_k \lambda^k
\end{equation}
($k$-is natural, and $A_k\neq 0$ - are complex constants), which is of
special interest.

Note that in paper [3] we have obtained the function $\Delta (\lambda )$ ($%
\Delta (\lambda )$ -is called a characteristic function) which is an entire
analytical function on complex parameter $\lambda $ and the studying a some
its properties (for example distribution of zeros, distance between two
zeros, lower estimation for modules and so on) is very important step in
spectral theory of differential operators.

The present paper is continuation of [3] and the function of the form (1) is
a special case\thinspace \thinspace of $\Delta (\lambda ).$

Infroduce into consideration the following sets of points of the complex
surfare $C:$

$$
\Omega _{R_1R_2}(\lambda _0)=\left\{\lambda ;R_1\leq \left| \lambda
-\lambda _0\right| \leq R_2\right\},\,\,\,\Omega _{R_1,R_2}=\,\,\Omega
_{R_1,R_2}(0),
$$
$$
\Omega _R(\lambda _0)=\left\{\lambda ;\left| \lambda -\lambda _0\right|
\leq R\right\} ,\,\,\,\Omega _R=\,\,\Omega _R(0),
$$
$$
\Gamma _{kj}^S(h,R)=\left\{\lambda;\,Re\lambda +(-1)^Sk\ln \left| \lambda
\right| =h,\,\,\,(-1)^jJm\lambda <0\right\} \cap \Omega _{R,\infty },
$$
$$
\Gamma _k^S(h,R)= \stackunder{j=1}{\stackrel{2}{U}}\Gamma _{kj}^S,
$$
$$
\Pi _{kj}^S(h,R)=\left\{ \lambda ;\,\left| Re\lambda +(-1)^Sk\ln \left|
\lambda \right| \right| \leq h,\,\,\,(-1)^jJm\lambda <0\right\} \cap \Omega
_{R,\infty },
$$
\begin{equation}
\label{2}\Pi _k^S(h,R)=\stackunder{j=1}{\stackrel{2}{U}}\Pi _{kj}^S(h,R),
\end{equation}

$$
T_{k1}^S(h,R)=\left\{ \lambda ;\,Re\lambda +(-1)^Sk\ln \left| \lambda
\right| <-h\right\} \cap \Omega _{R,\infty },
$$
$$
T_{k2}^S(h,R)=C\backslash T_{k1}^S(h,R)\cup \Pi _k^S(h,R)\cup \overline{%
\Omega }_{0,R},
$$
$$
T_k^S(h,R)=\stackunder{j=1}{\stackrel{2}{U}}T_{kj}^S(h,R),
$$
$$
\Sigma _\delta ^{(i)}=\left\{ \lambda ;\,\,\left| \arg \lambda +(-1)^i\frac
\pi 2\right| <\delta \right\} ,
$$
where $0\leq R_1<R_2<\infty,\,\,\,R>0,\,\,\,h>0,\,\,\,i=1,2;\,\,j=1,2;\,\, \\
S=1,2;\,\,\,\delta >0$

Now let's investigate of some properties of the function $f_k(\lambda)$:

$$
\left| f_k(\lambda )\right| \geq \left| A_k\right| \left| \lambda \right|
^k\left[ 1-\left| B_k\right| e^{Re\lambda -k\ln \left| \lambda \right|
}\right] \geq \left| A_k\right| \left| \lambda \right| ^k\left[ 1-\left|
B_k\right| e^{-h}\right] ,
$$
where $\left| B_k\right| =\frac 1{\left| A_k\right| }$

At the condition $\lambda \in T_{k1}^1(h,R)$ and choosing $h>\ln 2\left|
B_k\right| $ we arrive at the estimation of the form
\begin{equation}
\label{3}\left| f_k(\lambda )\right| \geq \frac 12\left| A_k\right| \left|
\lambda \right| ^k,
\end{equation}
and at the condition $\lambda \in T_{k2}^2(h,R)$ we arrive at the following
estimation of the form
\begin{equation}
\label{4}\left| f_k(\lambda )\right| \geq \frac 12\left| e^\lambda \right|
,\,\,\text{if}h>\ln 2\left|A_k\right|
\end{equation}

Thus it was proved:

Lemma 1. The function $f_k(\lambda )$ at the sufficiently large $h>0,\,\,R>0$
in the domain $T_{k1}^1(h,R)$ and $T_{k2}^2(h,R)$ hasnt zeros. And what is
more the estimations (3) and (4) are true for it. Absence of zeros of the
function $f_k(\lambda )$ in the domain $T_{k1}^1(h,R)$ and $T_{k2}^2(h,R)$
is obvions from the estimations (3) and (4). Proceeding from the definition
of the curvilinear bands $\Pi _{kj}^S(h,R)$ the folloving is easily proved.

Lemma 2. For any $\delta >0$ and $h>0$ we can find $R>0$, such that

$$
\Pi _{kj}^S(h,R)\subset \Sigma _\delta ^{(1)}\cup \Sigma _\delta ^{(2)}
$$

Let's prove now the following lemma.

Lemma 3. The function $f_k(\lambda )$ in the complex surfare C has
denumerable sets of zeros $\left\{ \lambda _{\nu k}\right\} $ with unique
limit point $\lambda =\infty $ which at sufficiently large $h>0,R>0$, is
situated in the domain $\Pi _k^1(h,R)\cup \overline{\Omega }_{0,R}$ . These
zeros allow the asymptotic representation:

\begin{equation}
\label{5}\lambda _{\nu k}=\ln \frac{\left| A_k\right| }{\left[ 2\pi \left|
\nu \right| \right] ^k}+i\left( 2\pi \nu +\pi +\frac{\pi k}2+\arg A_k\right)
+0\left( \frac{\ln \left| \nu \right| }\nu \right)
\end{equation}

Proof. The assertion of the first part of the lemma follows from the general
theory of Picard [4] and from lemma 1. Prove the second part of the lemma.

$$
f_k(\lambda )=0
$$
$$
e^\lambda +A_k\lambda ^k=0
$$
$$
e^\lambda \cdot \lambda ^{-k}=-A_k
$$
$$
e^\lambda \cdot e^{-k\ln \lambda }=-A_k
$$
$$
\lambda -k\ln \left| \lambda \right| =\ln \left| A_k\right| +i\left( \arg
(-A_k)+2\pi \nu \right) .
$$

make the substitution $\lambda -2\pi \nu i=\xi _\nu ,$

then

$$
\xi _\nu =\ln \left| A_k\right| +i\arg (-A_k)+k\ln \lambda =
$$
$$
=\ln \left| A_k\right| +i\left( \arg A_k+\pi \right) +k\ln \left( 2\pi \nu
i+\xi _\nu \right) =
$$
$$
=\ln \left| A_k\right| +i\left( \arg A_k+\pi \right) +k\ln \left[ 2\pi \nu
i\left( 1+\frac{\xi _\nu }{2\pi \nu i}\right) \right] =
$$
$$
=\ln \left| A_k\right| +i\left( \arg A_k+\pi \right) +k\ln (2\pi \nu i)+k\ln
\left( 1+\frac{\xi _\nu }{2\pi \nu i}\right) .
$$
since%
$$
\stackunder{\nu \rightarrow \infty }{\lim }\ln \left( 1+\frac{\xi _\nu }{%
2\pi \nu i}\right) =0,\,\ln \,\,\left( 1+\frac{\xi _\nu }{2\pi \nu i}\right)
=0\left( \frac{\xi _\nu }{2\pi \nu i}\right) =0\left( \frac{\ln \left| \nu
\right| }\nu \right) .
$$
$$
\xi _\nu =\ln \left| A_k\right| +i\left( \arg A_k+\pi \right) +k\ln (2\pi
\nu i)+0\left( \frac{\ln \left| \nu \right| }\nu \right) ,
$$
then we find
$$
\lambda _\nu =\ln \frac{\left| A_k\right| }{\left[ 2\pi \left| \nu \right|
\right] ^{-k}}+i\left( 2\pi \nu +\pi +\frac{k\pi }2+\arg A_k\right) +0\left(
\frac{\ln \left| \nu \right| }\nu \right) .
$$

From (5) we see that at large $\left| \nu \right| $ we have

\begin{equation}
\label{6}\left| \lambda _{\nu +1}-\lambda _\nu \right| =2\pi +0(1)
\end{equation}

Consequently there exists $\delta >0$ such that the circles of $\Omega
_\delta (\lambda _\nu )$ are mutually exclusive and at the sufficiently large
$h>0,R>0$ wholly lie in the domain $\Pi _k^1(h,R)\cup \overline{\Omega}%
_{0,R}$ . From the asymptotic formulae (5) and (6) we see that
straight lines
$$
l_{\nu ,k}=\left\{ \lambda :\,\,Jm\lambda =Jm\lambda _\nu -\left( \pi +\frac{%
\pi k}2+\arg A_k\right) \right\}
$$
are perpendicular to the imaginary axis $Re\lambda =0$ at all possible
different, sufficiently large (by the modulus) values $\nu $ are different
and divide the domains $\Pi _k^1(h,R)$ into the curvilinear quadrangles $%
D_{\nu k}=D_{\nu k}(h,R)$ with lateral boundaries on the lines $\gamma
_k(-h,R),\,\gamma _k(h,R)$ and with the base on the straight lines $l_{\nu
-1,k}>l_{\nu ,k}$.The length of the diogonal of a quadrange denote by
$$
d_\nu =\stackunder{\lambda ,\mu \in D_{\nu ,k}}{\sup }\left| \lambda -\mu
\right| .
$$
Introduce the folloving notation $\Pi _k^1(h,R,\delta )=$ $\Pi
_k^1(h,R)\setminus \stackunder{\nu }{U}\Omega _{0,\delta }(\lambda _\nu )$,
where the sigh of unification is propagated on all $\nu $ such that

$$
\lambda _\nu \in \Pi _k^1(h,R),
$$
$$
D_{k\nu }^\delta =D_{k\nu }\setminus \Omega _{0,\delta }(\lambda _\nu ).
$$
The following lemma is true.

Lemma 4. There exists the constant $\delta >0$ , such that at $\lambda \in
\Pi _k^1(h,R,\delta )$ it halds the inequality

\begin{equation}
\label{7}\left| f_k(\lambda )\right| \geq C_\delta \left| \lambda \right| ^k
\end{equation}


\begin{thebibliography}{99}
\bibitem{1}  Levin B.Ya., Distribution of roots of entire functions.
Gostekhizdat, Moscow, 1956.
\bibitem{2}  Sadovnichii V.A., Lyubushkin V.A., Belabbasi Yu., On a
zeros of a class of whole functions.In the book:''Proceedings of the
I.G.Petrovskiy's workshops'',1982, p.211-217
\bibitem{3} Mamedov Yu.A., Ahmadov H.I., Almost Regularity conditions
of Spectral problems for a Second order Equations. Math. arxiv:
math-ph/0212073.
\bibitem{4} Privalov I.I., Introduction to the theory of complex
variable function, ''Nauka'', Moscow, 1977, 444p.
\bibitem{5} Rasulov M.L., Methods of contour integration. - Amsterdam:
North Holland publishing company, 1967, 462p.
\end{thebibliography}
\end{document}